\documentclass[jjapprn]{jjap2}
\title{Effect of Energetic Disorder on the Open-Circuit Voltage in Organic Bulk Heterojunction Composites}
\author{Kazuhiko~{\sc Seki}$^1$, Kazuhiro~{\sc Marumoto}$^2$, and Masanori~{\sc Tachiya}$^1$}
\inst{$^1$National Institute of Advanced Industrial Science and Technology (AIST),
Tsukuba Ibaraki 305-8565, Japan\\
$^2$ Division of Materials Science, University of Tsukuba, Tsukuba, Ibaraki 305-8573, Japan
}
\abst{
Under open-circuit condition, 
the current is not extracted and 
the photogenerated carriers in principle disappear only by recombination.
We study the open-circuit voltage $V_{\rm OC}$ and transient photovoltage 
under the effect of bulk recombination in a medium with energetic disorder  
by using the multiple trapping (MT) model.
The key parameter in the MT model is the dispersion parameter $\alpha$ given by 
the ratio of thermal energy to the characteristic energy of trap states.  
We show that $V_{\rm OC}$ depends linearly on the logarithm of the light intensity and 
the slope depends on the $\alpha$ of the MT model.  
Under the continuous irradiation of light, 
the photovoltage response to the weak perturbation by a pulsed light 
obeys pseudo-first-order decay. 
The rate as a function of $V_{\rm OC}$ is 
independent of the dispersion parameter. 
However, it obeys the power law as a function of light intensity,  and 
the exponent is given by 
$1/(1+\alpha)$, which reduces to $1/2$ in the absence of energetic disorder.  
}
\begin{document}
\maketitle
\sloppy
\section{Introduction}
One of the factors limiting 
the efficiency of organic photovoltaic cells is  the charge recombination between strongly bound positive and negative charges.~\cite{Deibel} 
Organic photovoltaic cells are composed of organic materials of low dielectric constants.  
As a result, 
positive and negative charges are attracted by the long-range 
Coulombic interaction 
of low-dielectric materials. 
The efficiency of organic photovoltaic cells is also closely related to structural disorder 
inherent in amorphous dielectric materials.~\cite{Ohkita2010} 
The commonly used bulk heterojunction structure contains microdomains of disordered phases  
as it is difficult to make a perfectly ordered blend of two-phase systems.  
The structural disorder in dielectric materials creates the distribution of electrostatic energy at charge trap sites,  
which is characterized by the trapping energy distribution. 
In this study, we examine the open-circuit voltage ($V_{\rm OC}$ )
under the bulk recombination incorporating charge trap states with 
the trapping energy distribution. 

The kinetics of charge recombination incorporating charge trap states with the  
trapping energy distribution was investigated by simulating  a 
thermally activated hopping model 
where charge recombination is limited
by the random walk of charges towards the counter charges;  
the random walk is mediated
by trapping and detrapping from a limited number of deep traps.~\cite{Nelson99,Nelson03} 
The model differs from the continuous time random walk model; 
it prevents the multiple occupancy of sites 
and requires intensive numerical simulations.~\cite{Nelson99,Nelson03} 
A simplified version of the model, amenable to analytical treatment, 
was studied to elucidate the kinetics of charge recombination processes.~\cite{Barzykin02,Barzykin04,Tachiya10} 
The model was based on the multiple trapping (MT) model,~\cite{Schmidlin,Rudenko1,Rudenko2,Rudenko3} 
where trapping and detrapping from a limited number of deep traps were fully taken into account but  
the spacial distribution of carriers was not taken into account.~\cite{Barzykin02,Barzykin04,Tachiya10} 
According to the random walk simulation or the MT model, 
the decay of the total density of charges after they were generated by a short pulse of the excitation light was slower than 
that predicted using an ordinary bimolecular reaction model.~\cite{Nelson99,Nelson03,Barzykin02,Barzykin04,Tachiya10} 
In the MT model, 
the decelerated decay of the total density of charges reflects the decrease in detrapping rate as 
the distribution  of charges among trap states shifts  toward deeper trap levels  
inside the band gap.~\cite{Barzykin02,Tachiya10} 
A similar feature was also revealed by other approaches.~\cite{Zaban,Foertig} 
Recently, several groups have observed the decelerated decay of carriers in bulk heterojunction composites.~\cite{Nogueira,Eng,Ohkita,Ohkita2008,Mozer,Montanari,Juska,ShuttlePRB,Clarke,Naito} 
The results can be interpreted using the MT model.~\cite{Tachiya10} 

The MT model can also be applied to the study of 
the steady state carrier distribution among trap sites with different trapping energies. 
The carrier distribution 
can be changed by changing the intensity of continuous light,  
and the detrapping rates are changed accordingly. 
By using the MT model, the variation in detrapping rates by changing the light intensity 
can be fully taken into account, and 
we have shown theoretically that the total density of carriers depends on the light intensity,  
as observed in the experiments.~\cite{Seki13}  

If the total density of carriers is changed in steady state experiments by changing the light intensity, 
it changes the distribution of carriers among trap sites with different trapping energies. 
Under open-circuit condition, 
the current is not extracted and 
the photogenerated carriers in principle disappear only by recombination.
When recombination is in equilibrium with the generation of charge pairs, 
$V_{\rm OC}$   
is given by the difference between the quasi-Fermi energy of a hole and 
that of an electron both measured from the same level.~\cite{Wurfel,Garcia-Belmonte}
This affects 
$V_{\rm OC}$. 
In the present work, 
we study the dependence of $V_{\rm OC}$ on light intensity 
using the MT model. 

Recently, 
the transient decay of photovoltage has been measured.~\cite{Shuttle,Li} 
The transient photovoltaic decay was measured by applying a low intensity of pulsed light 
under the irradiation of strong continuous light. 
Pseudo-first-order decay has been typically observed. 
In this work, 
we show that the decay rate as a function of continuous light intensity is affected by the trapping energy distribution,  
but that as a function of  $V_{\rm OC}$ is insensitive to the exponential trapping energy distribution . 
To corroborate our theory, 
we also calculate the decay of  $V_{\rm OC}$ after the continuous light is completely turned off.

For simplicity, we assume that  
the kinetic parameters of electrons and holes are identical. 
Previously, we assumed that the detrapping frequencies of holes and electrons are very different.~\cite{Tachiya10,Seki13} 
When both electrons and holes are equally mobile among trap sites obeying the same trapping energy distribution, 
the former approach is relevant in the study of $V_{\rm OC}$. 

In Sect. 2, $V_{\rm OC}$ is formulated. 
In Sect. 3, $V_{\rm OC}$ in the steady state is calculated using 
the MT model when the kinetic parameters of electrons and holes are the same. 
In Sect. 4, the transient kinetics is studied 
by using the MT model. 
The last section is devoted to the conclusion. 

\section{Open-Circuit Voltage}
\begin{figure}
\centerline{
\includegraphics[width=0.5\columnwidth]{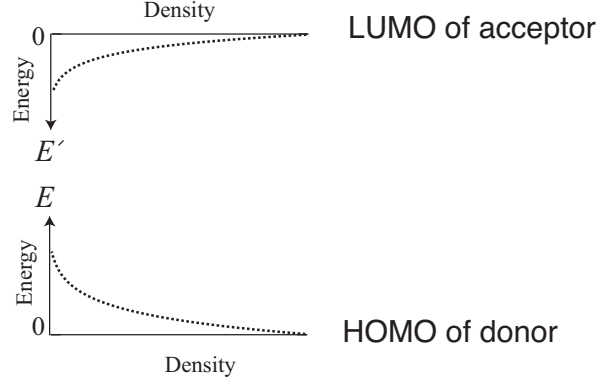}
}
\caption{Schematic representation of $E$ and $E'$ coordinates of trap states. The corresponding dimensionless energies are denoted by   
$\epsilon=E/(k_BT)$ and $\epsilon'=E'/(k_BT)$, respectively.} 
\label{fig:polaronenergy}
\end{figure}

The distribution of holes among trap states with different trapping energies can be characterized by 
the quasi-Fermi energy. 
We define the quasi-Fermi energy of a hole, $E_F$, from the highest occupied molecular orbital (HOMO) level of the donor in the upward direction 
(see Fig. \ref{fig:polaronenergy}).  
The open-circuit voltage  
is given by the difference between the quasi-Fermi energy of a hole and 
that of an electron both measured from the same level, for example, the HOMO level of the donor is,~\cite{Wurfel,Garcia-Belmonte}
\begin{align}
eV_{\rm OC}=  E_F^{(e)} - E_F   ,   
\label{Voc_fermi0}
\end{align}
where $E_F^{(e)}$ is the quasi-Fermi level of an electron measured from the HOMO level of the donor. 

The quasi-Fermi energy of an electron, $E_F'$, is measured from the 
lowest unoccupied molecular orbital (LUMO) level of the acceptor 
in the downward direction (see Fig. \ref{fig:polaronenergy} for the energy coordinates). 
We consider the case when the kinetic parameters of electrons and holes are the same. 
We also assume that  
the trapping energy distributions of electrons and holes are the same.  
In this case, 
the quasi-Fermi energy of an electron measured from the HOMO level of the donor 
in the upward direction is given by  
\begin{align}
E_F^{(e)}&= \mbox{acceptor LUMO level} - \mbox{donor HOMO level} -E_F ,  
\label{eq:28_0} 
\end{align}
where $E_F={E}_F'$ is used. 
Equation (\ref{Voc_fermi0}) can be expressed 
using Eq. (\ref{eq:28_0}) as  
\begin{align}
eV_{\rm OC}= \mbox{acceptor LUMO level} - \mbox{donor HOMO level} - 2E_F   .    
\label{Voc_fermi}
\end{align}

\section{Open-Circuit Voltage in Steady State }
In the steady state, 
the quasi-Fermi energy depends on the light intensity according to the 
distribution of holes among trap states with different trapping energies changed by varying the intensity of continuous light. 
This implies that the open-circuit voltage 
estimated from the quasi-Fermi energy is affected by the  light intensity. 

By light irradiation, excitons are generated. 
The excitons at the interface between the donor and  acceptor phases can be dissociated into 
a hole and an electron. 
The charge generation rate $G$ is proportional to the light intensity $I$, 
\begin{align}
G=k_g I,  
\label{eq:GvsI}
\end{align} 
where $k_g$ is the proportional coefficient. 
In the MT model, both holes and electrons are trapped in the trap states. 
By assuming that the number density of trap sites for electrons is the same as that for holes, 
the number density of trap sites can be denoted by $N$ for both electrons and holes. 
The total number density of holes, $n$, is the same as that of electrons since  
charges are generated and recombine with pairs.  
We introduce the trapping energy $E$ of a hole measured relative to the HOMO
of the donor (see Fig. \ref{fig:polaronenergy}). 
The detrapping rate constant is given by 
\begin{align}
k_d(E) = \nu_d \exp \left(\frac{-E}{k_B T} \right),                                  
\label{kd}
\end{align}
where $\nu_d$ is the detrapping frequency, $k_B$ the Boltzmann constant, and 
$T$ the temperature. 

The trapping energy distribution in organic materials including components 
of bulk-heterojunction solar cells can be represented by 
an exponential distribution.~\cite{Beiley,Mark,Blom,Campbell,Chiguvare,Street}
The exponential distribution of trapping energies, $g(E)$, can be represented by  
\begin{align}
g(E) = \frac{1}{E_0}\exp\left(-\frac{E}{E_0} \right),  
\label{gE}                            
\end{align}
where $E_0$ is a parameter indicating the characteristic trapping energy.  

By thermal excitation, holes can be detrapped to the free state. 
The holes in the free state can either be trapped by vacant trap sites 
at the rate $k_t$ or recombine with trapped electrons at the rate $k_r$. 
For simplicity, we assume that the kinetic parameters of electrons are the same as those of holes 
and that the trapping energy distribution of electrons obeys the same exponential form. 
Therefore, we use the same notations for kinetic parameters and $E_0$ for electrons. 

$\nu_d$ will be used to normalize rates 
and dimensionless energy will be introduced using $\epsilon = E/k_B T$. 
In the case when we need to distinguish the $\epsilon$ of electrons from that of holes, 
we denote it by $\epsilon'$ (see also Fig. \ref{fig:polaronenergy}). 
$\epsilon'$ is measured from the LUMO level of the acceptor 
in the downward direction. 
We denote the number density of holes trapped at trap sites with the trapping energy $\epsilon$ 
by $f(\epsilon)$. 
The notation  
$f(\epsilon)$ will also be used for electrons since the kinetic parameters of electrons are the same as those of holes.  

In the dimensionless unit, the trapping energy distribution is expressed as  
\begin{align}
g(\epsilon) &= \alpha \exp(-\alpha \epsilon) .     
\label{galpha}      
\end{align}
$\alpha = k_B T/E_0$ is a key quantity in the MT model and is called the dispersion parameter. 
The dispersion parameter is the ratio of thermal energy to the characteristic energy of trap states. 
When thermal energy is lower than the typical trapping energy of trap states, 
$\alpha$ is smaller than unity.  
When $\alpha$ is small, detrapping assisted by thermal energy is suppressed.

The holes initially produced in the free state can be trapped 
into a trap state with the trapping energy $\epsilon$ with the probability proportional to $k_t$ 
multiplied by the number density of the vacant trap sites for holes given by  
$N g(\epsilon) - f(\epsilon)$. 
The probability that a hole in the free state will recombine with a trapped electron is proportional to 
the recombination rate multiplied by the total number density of electrons given by 
\begin{align}
n= \int_0^\infty d \epsilon f (\epsilon). 
\label{totaln}
\end{align}
The probability that a hole in the free state will be trapped into a site with the trapping energy $\epsilon$ 
becomes~\cite{Tachiya10,Seki13} 
\begin{align} 
\frac{k_t[Ng(\epsilon) - f(\epsilon)]}{k_t N + (k_r - k_t)n} ,  
\label{expl1}
\end{align}
where $ N -  n = \int_0^\infty d \epsilon [Ng(\epsilon) - f(\epsilon)]$ is used. 
This factor is multiplied by the hole 
generation rate in the free state, $G$, and 
the number density of holes detrapped into the free state from all traps per unit time,  
$\int_0^\infty e^{-\epsilon} f(\epsilon) d \epsilon$, 
to obtain the growth rate of holes with the trapping energy $\epsilon$, 
\begin{align}
\left( G + \int_0^\infty e^{-\epsilon} f(\epsilon) d \epsilon\right) \frac{k_t \left[ N g (\epsilon) -f(\epsilon) \right]}{k_t (N-n) + k_r n}
.   
\label{eq:growth}
\end{align}

Electrons produced in the free state can recombine with holes and the probability per unit time is 
given by 
\begin{align}
\frac{k_r f(\epsilon)}{k_t N + (k_r - k_t)n} . 
\label{expl1_1}
\end{align}
This factor is multiplied by the electron 
generation rate in the free state, $G$, and 
the number density of electrons detrapped into the free state from all traps per unit time,  
$\int_0^\infty e^{-\epsilon} f(\epsilon) d \epsilon$, 
to obtain the annihilation rate of holes with the trapping energy $\epsilon$ attributable to recombination, 
\begin{align}
\left( G + \int_0^\infty e^{-\epsilon} f(\epsilon) d \epsilon\right) \frac{k_r f (\epsilon)}{k_t (N-n) + k_r n}
.   
\label{eq:annihilation}
\end{align}
In the steady state, holes with the trapping energy $\epsilon$ 
generated by trapping should be 
balanced with those 
annihilated   
by recombination and detrapping, 
\begin{align}
0= \left( G + \int_0^\infty e^{-\epsilon} f(\epsilon) d \epsilon\right) \frac{k_t \left[ N g (\epsilon) -f(\epsilon) \right]-
k_r f (\epsilon)}{k_t (N-n) + k_r n}- e^{-\epsilon} f(\epsilon)
,   
\label{appendixB_1}
\end{align}
where $e^{-\epsilon} f(\epsilon)$ is the detrapping rate of a hole from a trap state with the trapping energy $\epsilon$. 
As already stated, all the rates including $e^{-\epsilon} f(\epsilon)$ are normalized by $\nu_d$. 

By integrating Eq. (\ref{appendixB_1}) over $\epsilon$, 
we obtain  
\begin{align}
G= \frac{2 k_r n \left(G+ \int_0^\infty e^{-\epsilon} f(\epsilon) d \epsilon \right)}{k_t (N-n) + k_r n} .
\label{appendixB_2}
\end{align}
Equation (\ref{appendixB_1}) can be rewritten by using Eq. (\ref{appendixB_2}) as 
\begin{align}
f(\epsilon) =\left( \frac{k_t}{k_t + k_r} \right) \frac{Ng(\epsilon)}{1+ \exp \left[- \left(\epsilon-\epsilon_F\right) \right] }, 
\label{appendixB_3}
\end{align}
where $\epsilon_F$ is given by  
\begin{align}
\epsilon_F= \ln \left( \frac{N}{G} \frac{2k_r}{k_t + k_r} \rho \right),  
\label{appendixB_4}
\end{align}
and $\rho$ represents the fraction of trap sites occupied by holes  
\begin{align}
\rho=n/N. 
\label{eq:rho}
\end{align}
Equation (\ref{appendixB_3}) represents the Fermi-Dirac distribution with the quasi-Fermi energy given by $\epsilon_F$. 
The quasi-Fermi energy is related to $E_F$ in Eq. (\ref{Voc_fermi0}) by $\epsilon_F = E_F/k_B T$.  
By introducing Eq. (\ref{appendixB_3}) into Eq. (\ref{totaln}), 
we obtain, using Eqs. (\ref{eq:rho})  and (\ref{galpha})  
\begin{align}
\rho = \frac{k_t}{k_t+k_r} 
\int_0^\infty d \epsilon 
\frac{g(\epsilon)}{1+\exp \left[- \left(\epsilon-\epsilon_F\right) \right]} .
\label{appendixB_4_1}
\end{align} 
By expanding the integrand in terms of $\exp \left( \epsilon_F \right)$, Eq. (\ref{appendixB_4_1}) can be rewritten as
\begin{align}
\rho &=\frac{k_t}{k_t+k_r} \sum_{j=0}^\infty (-1)^j\exp \left( j \epsilon_F \right) \hat{g} (j),
\label{expandrhos_1}
\end{align}
where $\hat{g} (j)$ denotes the Laplace transform of $g(\epsilon)$, 
$\hat{g} (j) = \int_0^\infty d \epsilon \exp (-j \epsilon) g(\epsilon)$. 
We note $\hat{g} (0)=1$ from the normalization condition.  
We obtain $\hat{g} (j) =\alpha/(j+ \alpha)$ for the exponential trapping energy distribution. 
By using this equation,  
Eq. (\ref{expandrhos_1}) can be expressed as  
\begin{align}
\rho= \frac{k_t}{k_t+k_r}  \mbox{}_2 F_1 
\left(1,\alpha,1+\alpha, -\exp \left( \epsilon_F \right) \right), 
\label{expandrhos}
\end{align}
where $\mbox{}_2 F_1 
\left(1,\alpha,1+\alpha, - x\right)$ is the hypergeometric function.~\cite{Abramowitz} 

\subsection{Partly trap filled regime}
We consider the case $\exp \left( \epsilon_F \right) \gg1$.
The condition is equivalent  to
\begin{align}
\frac{\displaystyle 2k_t k_r}{\displaystyle (k_t+k_r)^2} \frac{N}{G} \gg1. 
\label{both_small}
\end{align} 
When this condition is satisfied, 
by utilizing the  asymptotic expansion of $\mbox{}_2 F_1 
\left(1,\alpha,1+\alpha, -x \right)$,~\cite{Abramowitz} 
we obtain 
\begin{align}
\rho \simeq \left( \frac{\pi \alpha}{\sin (\pi \alpha) } 
\frac{k_t}{k_t+ k_r} \right)^{1/(\alpha+1)} 
\left(\frac{k_t+ k_r}{2 k_r}\frac{G}{N} 
\right)^{\alpha/(\alpha+1)}.   
\label{appendixB_5}
\end{align}
Equation (\ref{appendixB_5}) shows that 
the number density of holes obeys the power law dependence on light intensity. 
By introducing Eq. (\ref{appendixB_5}) into Eq. (\ref{appendixB_4}), 
the quasi-Fermi energy of a hole given by 
Eq. (\ref{appendixB_4}) is expressed as 
\begin{align}
\epsilon_F = \frac{1}{\alpha + 1} \left[
\ln \left(\frac{\pi \alpha}{\sin \pi \alpha} \right) + 
\ln \left(\frac{2 k_t k_r}{(k_t+k_r)^2} \right)  
- \ln (G/N) 
\right]. 
\label{appendixB_6}
\end{align}
Essentially the same equation as Eq. (\ref{appendixB_5}) has been obtained by assuming that  
the detrapping frequencies of holes and electrons are very different.~\cite{Seki13} 
Here, we show that power law dependence of hole density on light intensity 
is given by the same exponent even when   
the detrapping frequencies of holes and electrons are the same. 

By substituting Eq. (\ref{appendixB_6}) into Eq. (\ref{Voc_fermi}) and considering $\epsilon_F = E_F/k_B T$,  
$V_{\rm OC}$ is obtained as  
\begin{align}
V_{\rm OC} = c_2 + \left(\frac{2}{\alpha+1} \right) \frac{k_B T}{e} \ln I , 
\label{VOC_pe}
\end{align}
where $c_2$ is a constant independent of light intensity. 
The slope of $V_{\rm OC}$ against $\ln I$ could be in the range between $k_B T/e$ and $2 k_B T/e$ in this case. 
A slope larger than $k_B T/e$ has recently been reported for bulk heterojunction solar cells, 
where charge transport is  
trap-limited.~\cite{Mandoc, MandocAPL2007,Cowan} 

Recently, 
the relations given by Eq. (\ref{VOC_pe}) have also been found by numerical simulation.~\cite{Blakesley}
In the simulation, 
the drift-diffusion equation was solved together with Poisson's equation, 
and the carrier density in the trap states, $\rho_t$, 
was approximated as $\rho_t \sim \exp \left(-E_f'/E_0\right)$ in terms of the 
quasi-electrochemical potential $E_f'$ determined self-consistently. 
The results were interpreted using $\rho_t$ given above with the 
spatially homogeneous $E_f'$.~\cite{Blakesley,Kirchartz}  
Here, we calculate $\rho$ and $V_{\rm OC}$ analytically without assuming their forms. 

\subsection{Completely trap filled regime}
Let us consider the limit $\exp \left( \epsilon_F \right)\ll1$ that is equivalent to  
\begin{align}
\frac{\displaystyle 2k_t k_r}{\displaystyle (k_t+k_r)^2} \frac{N}{G} \ll1. 
\label{trapfilling}
\end{align}
This condition is satisfied at a high light intensity leading to the high carrier generation rate $G$. 
Under this condition, we obtain from Eq. (\ref{expandrhos_1})  
\begin{align}
\rho \simeq  
\left( \frac{k_t + k_r}{k_t }+\frac{2 k_r \hat{g} (1)}{k_t+ k_r} \frac{N}{G} \right)^{-1} . 
\label{saturho}
\end{align}
In this limit, the number density saturates as the generation rate $G$ increases.  
The fraction $\rho$ of trap sites occupied by carriers saturates when the light intensity is high. 
The saturation results from the trap filling 
of the deep trap levels inside the band gap up to the shallow trap levels. 
Note that Eq.  (\ref{saturho}) is derived without assuming any form of trapping energy distribution.   
$\hat{g} (1)=\alpha/(1+\alpha)$ is obtained for the exponential trapping energy distribution. 
By using Eq. (\ref{saturho}), Eq. (\ref{appendixB_4}) can be expressed as  
\begin{align}
\exp \left( \epsilon_F \right) &\simeq 
\left[ 
\hat{g}(1) + \frac{(k_t+k_r)^2}{2k_tk_r} \frac{G}{N}  
\right]^{-1} 
\nonumber \\
&\simeq \frac{2 k_t k_r}{(k_t+k_r)^2} \frac{N}{G}, 
\label{ysatG}
\end{align}
where Eq. (\ref{trapfilling}) and $1> \hat{g} (1)$ are used. 
As a result, $\epsilon_F $ is obtained as 
\begin{align} 
\epsilon_F \simeq \ln \left( \frac{2 k_t k_r}{(k_t+k_r)^2}\right) - \ln (G/N) .
\label{eFtrapfilling}
\end{align}
By substituting Eq. (\ref{eFtrapfilling}) with Eq. (\ref{eq:GvsI}) in Eq. (\ref{Voc_fermi}), 
$V_{\rm OC}$ is obtained using $\epsilon_F = E_F/k_B T$ as a linear function of the logarithm of light intensity, $I$, 
\begin{align}
V_{\rm OC} = c_1 + 2(k_B T/e) \ln I.   
\label{VOC_trapfilling}
\end{align}
Note that when the light intensity is high, this relation   
is derived without assuming any specific forms of $g(\epsilon)$. 
The slope of $V_{\rm OC}$ vs $\ln I$ is given by $2 k_B T/e$ 
regardless of the form of $g(\epsilon)$ as a result of the trap filling effect. 
Equation (\ref{VOC_trapfilling}) is the same as that 
known for  a single trap state both for a hole and 
an electron. \cite{Koster05,Cowan,Deibel10} 
When the light intensity is high, shallow trap levels close to the conduction band are also occupied. 
Charges occupying in shallow trap states are easily detrapped to the conduction band;  
thus, they behave similarly to quasifree charges. 
Accordingly, $\rho$ saturates when the light intensity is high.
The trap filling effect is important when the light intensity is high 
and the number of trap states is small. 

\section{Transient Photovoltage Decay}
\subsection{Under continuous light irradiation}
Recently, the transient decay of photovoltage induced by a small perturbative light pulse has been measured 
under open-circuit condition and continuous light irradiation. 
By pulsed irradiation, 
the charge density is increased from that generated by continuous light irradiation. 
Under the condition that a sufficiently small number of charges is generated by a light pulse compared with that generated by continuous light, 
the pseudo-first-order decay of the perturbed density can be obtained. 
In this section, we study  the decay rate as a function of light intensity 
and the effect of the exponential trapping energy distribution on this relationship. 
The transient photovoltage is obtained using Eq. (\ref{Voc_fermi}) with the time-dependent quasi-Fermi energy 
determined from the perturbed charge density. 

After light pulse irradiation at time $0$,
the carriers decay to the steady state, Eq. (\ref{appendixB_3}),  with the quasi-Fermi energy given by 
Eq. (\ref{appendixB_4}). 
The quasi-Fermi energy is expressed using the charge generation rate $G$ under the continuous light irradiation. 
The equation describing the transient decay of carriers is obtained 
by generalizing Eq. (\ref{appendixB_1}) to include 
the time dependence as   
\begin{align}
\frac{\partial f(\epsilon,\tau)}{\partial \tau} =
\left( G+ \int_0^\infty e^{-\epsilon} f(\epsilon,\tau) d \epsilon
\right)
\frac{k_t \left[ N g (\epsilon) -f(\epsilon,\tau) \right]-
k_r f (\epsilon,\tau)}{k_t (N-n) + k_r n}  - e^{-\epsilon} f(\epsilon,\tau)
.  
\label{eq:T1}
\end{align}
Following Ref.~\citeonline{Tachiya10}, 
we integrate Eq. (\ref{eq:T1}) over $\epsilon$ and obtain 
\begin{align}
\frac{\partial}{\partial \tau} \rho(\tau) =\frac{k_t - (k_r + k_t)\rho(\tau)}{k_t + (k_r - k_t)\rho(\tau)}  \frac{G}{N}-
\frac{2 k_r[\rho(\tau)]^2 \Phi(\tau)}{k_t + (k_r - k_t)\rho(\tau)} , 
\label{eqdensity2}
\end{align}
where
\begin{align}
\Phi(\tau)=\int_0^\infty d \epsilon \exp(-\epsilon) f(\epsilon, \tau)/n(\tau)   .
\label{defPhi}
\end{align}
One can calculate the decay kinetics of holes by solving Eqs. (\ref{eqdensity2}) and (\ref{defPhi}). 

We consider the case 
in which the recombination rate constant is much smaller  
than the trapping rate constant. 
In organic solar cells, 
recombination should be suppressed and 
this case is practically important. 
If the intrinsic recombination rate constant $k_r$ is much smaller 
than the trapping rate constant $k_t$, 
the effect of recombination on the distribution of holes among trap sites 
with different trapping energies is small. 
The distribution $f(\epsilon, \tau)$  is 
well approximated by a Fermi distribution with a quasi-Fermi energy 
$\epsilon_F(\rho)$, which depends on the fraction of trap sites occupied by holes, $\rho(\tau)$,  
\begin{align}
f(\epsilon, \tau) = \frac{Ng(\epsilon)}{\exp\left\{-[\epsilon-\epsilon_F(\rho(\tau))]\right\} +1} .
\label{Fermif}
\end{align}
By using the above equation, the factor 
$\int_0^\infty \exp(-\epsilon) f(\epsilon, \tau)d\epsilon$ in Eq. (\ref{defPhi}) is expressed as~\cite{Tachiya10}
\begin{align}
\int_0^\infty d\epsilon \exp \left( -\epsilon\right) f(\epsilon, \tau)/N  
&=  \exp[-\epsilon_F(\rho(\tau))]  \left( 1-\rho(\tau)\right). 
\label{Fermiphi}
\end{align}
By substituting the above equation, Eq. (\ref{eqdensity2}) becomes 
\begin{align}
\frac{d}{d\tau}\rho(\tau) =\frac{k_t - (k_r + k_t)\rho(\tau)}{k_t + (k_r - k_t)\rho(\tau)}\frac{G}{N}-2 k_r \exp[-\epsilon_F(\rho(\tau))] 
\frac{\rho(\tau)(1-\rho(\tau))}{k_t(1 - \rho(\tau)) + k_r \rho(\tau)} . 
\label{kineticFermi}
\end{align}
If $k_r$ is much smaller than $k_t$ and if $1 -\rho$ is not smaller than $\rho$, 
the above equation is approximated as
\begin{align}
\frac{d}{d\tau}\rho(\tau)  \approx (G/N) -k[\rho(\tau)]\rho(\tau),
\label{kineticFermiappr}
\end{align}
where the apparent rate constant $k[\rho]$ is given by  
\begin{align}
k[\rho] = 2(k_r/k_t)\exp[-\epsilon_F(\rho)] .
\label{knFermi}
\end{align}

The apparent rate constant $k[\rho]$ depends on the fraction of trap sites occupied by holes 
through the Fermi energy. 
The relationship between the quasi-Fermi energy and the fraction of trap sites occupied by holes is given by
\begin{align}
\int_0^\infty d\epsilon \frac{g(\epsilon)}{\exp[-(\epsilon-\epsilon_F(\rho))] +1} 
= \rho   .
\label{relationFermi}
\end{align}
As in Eqs. (\ref{expandrhos}) and (\ref{appendixB_5}), 
the left-hand side of Eq. (\ref{relationFermi}) can be expressed 
in terms of the hypergeometric function and can be approximated as
\begin{align}
\frac{\pi \alpha}{\sin(\pi \alpha)} \exp[-\alpha \epsilon_F(\rho)] = \rho ,  
\label{relationFermi1}
\end{align}
from which we obtain
\begin{align}
\epsilon_F(\rho)= - \frac{1}{\alpha} \ln \left(\rho \frac{\sin(\pi \alpha)}{\pi \alpha}\right) .
\label{relationFermi2}
\end{align}
The substitution of the above equation in Eq. (\ref{knFermi}) yields 
\begin{align}
k[\rho]  =2 (k_r/k_t)\left(\rho \frac{\sin(\pi \alpha)}{\pi \alpha}\right)^{1/\alpha}.  
\label{knFermi1}
\end{align}
According to Eq. (\ref{knFermi1}), the apparent rate constant is proportional to 
the $1/\alpha$-th power of the fraction of trap sites occupied by holes. 
The power law dependence with the unique exponent arises 
because the detrapping rate depends on the quasi-Fermi energy 
which in turn changes with the fraction of trap sites occupied by holes. 

We consider a small perturbation $\delta \rho$ from $\rho_0$ under continuous light irradiation. 
\begin{align}
\rho(\tau) =\delta  \rho(\tau) + \rho_0,
\label{eq:deltarho}
\end{align}
where $\rho_0$ is the steady state charge density obtained in the previous section, Eq. (\ref{appendixB_5}). 
When $\delta \rho \ll \rho_0$, 
the solution of Eq. (\ref{kineticFermiappr}) with Eq. (\ref{knFermi1}) is obtained as  
\begin{align}
\delta \rho(\tau) = \delta \rho(0) \exp \left( - k_{PV} \tau \right),
\label{eq:pseudofirst}
\end{align}
where the pseudo-first-order rate constant is 
\begin{align}
k_{PV}= \frac{\alpha + 1}{\alpha} \left( \frac{2 k_r}{k_t} \right)^{\alpha/(\alpha+1)} 
\left(
\frac{\sin (\pi \alpha)}{\pi \alpha}  
\right)^{1/(\alpha+1)} 
\left( \frac{G}{N} \right)^{1/(\alpha + 1)} . 
\label{eq:pseudofirstrate}
\end{align}
By using Eq. (\ref{relationFermi2}), the quasi-Fermi energy as a function of time becomes  
\begin{align}
\epsilon_F &= - \frac{1}{\alpha} \left\{ \ln \left(\rho_0 \frac{\sin(\pi \alpha)}{\pi \alpha}\right) + \ln\left(1+\frac{\delta \rho(\tau)}{\rho_0}\right) \right\}, 
\nonumber \\
& \simeq -c_F - \frac{1}{\alpha} \frac{\delta \rho(0)}{\rho_0} \exp \left( - k_{PV} \tau \right) , 
\label{eq:quasiFermit}
\end{align}
where $c_F$ is a constant independent of time. 
By combining Eqs. (\ref{eq:quasiFermit}) and (\ref{Voc_fermi}), we obtain the following using $\epsilon_F = E_F/k_B T$:  
\begin{align}
V_{\rm OC} (\tau) &= c_0  +\frac{2}{\alpha}\left(\frac{k_B T}{e}\right)\frac{ \delta \rho(0)}{\rho_0 } \exp \left( - k_{PV} \tau \right)   
\label{VOC_ts}
\end{align}
where $c_0$ is a constant independent of time. 
The transient photovoltage follows an exponential decay and the rate is given by Eq. (\ref{eq:pseudofirstrate}). 

When the charge generation rate $G$ is proportional to the light intensity $I$, 
the decay rate of transient photovoltage is expressed as 
\begin{align}
\ln k_{PV} \sim c_{I}+[1/(\alpha+1)] \ln I,  
\label{eq:pvvsI}
\end{align}
where $c_{I}$ is a constant independent of light intensity. 
The decay rate as a function of light intensity depends on the dispersion parameter.
By using
Eq. (\ref{VOC_pe}), Eq. (\ref{eq:pvvsI}) can be rewritten as  
\begin{align}
\ln k_{PV} \sim c_{V}+\left[e/(2 k_B T)\right] V_{\rm OC},  
\label{eq:pvvsVOC}
\end{align}
where $c_{V}$ is a constant independent of light intensity. 
The decay rate as a function of $V_{\rm OC}$ is independent of the dispersion parameter in sharp contrast to 
the expression of the light intensity dependence.  

\subsection{Under the dark}
Before closing this section, 
we consider the opposite limit which is the decay of the fraction of trap sites occupied by holes in the dark. 
The solution of Eq. (\ref{kineticFermiappr}) with Eq. (\ref{knFermi1}) is obtained as~\cite{Tachiya10}
\begin{align}
\rho = \frac{\rho_0}{[1 + (1/\alpha)[\rho_0 \sin(\pi \alpha)/(\pi \alpha)]^{1/\alpha}(2k_r /k_t)\tau]^{\alpha}} . 
\label{solFermi1}
\end{align}
For long times, $\rho(\tau)$ is further approximated as
\begin{align}
\rho \simeq \frac{\pi \alpha}{\sin(\pi \alpha)} \left(\frac{\alpha k_t }{2k_r}\right)^\alpha \tau^{-\alpha} .  
\label{solFermiapprox}
\end{align}
The recombination kinetics in this case exhibits dispersive kinetics. 
Essentially the same equations as Eqs. (\ref{solFermi1}) and (\ref{solFermiapprox}) have been obtained by assuming that  
the detrapping frequencies of holes and electrons are different.~\cite{Tachiya10} 
Here, we show that the power law dependence of hole density on time  
is given by the same exponent even when   
the detrapping frequencies of holes and electrons are the same.

Finally, we consider the decay of $V_{\rm OC}$ in the dark. 
By combining Eqs. (\ref{relationFermi2}) and (\ref{Voc_fermi}), we obtain the following using $\epsilon_F = E_F/k_B T$:  
\begin{align}
V_{\rm OC} (\tau) &= c_3  -2 \ln \left[ 1 + \frac{1}{\alpha}
\left(\rho_0 \frac{\sin(\pi \alpha)}{\pi \alpha} \right)^{1/\alpha} \frac{2k_r }{k_t}\tau \right] 
\label{VOC_t1}
\\
&\simeq c_4  -2 \ln \left( \tau \right),
\label{VOC_t2}
\end{align}
where $c_3$ and $c_4$ are constants independent of time. 
Equation (\ref{solFermi1}) is introduced to obtain the first equality and 
Eq. (\ref{solFermiapprox}) is introduced to obtain the second equality. 
The decay of the open-circuit voltage is not affected by the dispersion parameter unlike 
the charge density. 
In this sense, 
the kinetic measurement of $V_{\rm OC}$ in the dark is not useful for extracting the trapping energy distribution of 
trap states at least for the exponential trapping energy distribution.

\section{Conclusions}
The MT model is a powerful theoretical model used to investigate analytically 
bulk recombination between electrons and holes in semiconductors that possess energetic disorder. 
The key parameter in the MT model is the dispersion parameter $\alpha$ given by 
the ratio of thermal energy to the characteristic energy of trap states.  
When $\alpha$ is small, detrapping supported by thermal energy is suppressed. 

We have studied the MT model in the case when kinetic parameters of holes and electrons are the same. 
Previously, the MT model was formulated in the case 
when the detrapping frequencies of holes and electrons are different.~\cite{Tachiya10,Seki13}
In our previous work, 
the power law dependence of carrier density on excitation light intensity was obtained and the exponent 
was related to the dispersion parameter of carriers with a high detrapping frequency.~\cite{Seki13}
In this work, the exponent is given by the same function using  the dispersion parameter $\alpha$  
when the detrapping frequencies of holes and electrons are the same.  
In the case of transient kinetics, the asymptotic decay of carriers is given by the power law as a function of time, and 
the exponent as a function of the dispersion parameter is again the same as that obtained by assuming that 
the detrapping frequencies of holes and electrons are different.~\cite{Tachiya10}
We may conclude that the exponents of power law as functions of light intensity and time are not affected by 
the ratio of the detrapping frequency of holes to that of electrons. 
The exponents are given in terms of the dispersion parameter of carriers with a high detrapping frequency if they are different. 

By using the MT model, 
we show that the open-circuit voltage $V_{\rm OC}$ depends linearly on the logarithm of light intensity and 
that the slope is in the range between $k_B T/e$ and $2 k_B T/e$ 
depending on the dispersion parameter.  

Under continuous light irradiation, 
the photovoltage response to the weak perturbation by a light pulse 
obeys pseudo-first-order decay. 
The logarithm of the rate as a function of  $V_{\rm OC}$ has a slope given by 
 $e/(2 k_B T)$, which is independent of the dispersion parameter. 
 At room temperature, 
 the slope is $19$ V$^{-1}$ when $V_{\rm OC}$ is expressed by voltage. 
 The measured slope of $16$ V$^{-1}$ is $20$\% lower than the above value.~\cite{Shuttle}
 The slight decrease in the slope may partly be attributed to the temperature larger than $300$ K. 
 
Although the decay rate as a function of  $V_{\rm OC}$ is independent of the dispersion parameter, 
the decay rate as a function of the intensity of continuous light depends on the dispersion parameter. 
The decay rate as a function of light intensity obeys the power law and the exponent is given by 
 $1/(1+\alpha)$. 
 By combining the steady state result of $V_{\rm OC}$ as a function of light intensity  
 and the result obtained by the kinetic measurement, 
the dispersion parameter can be determined consistently. 
 By further changing the temperature, 
we can study the linearity of the dispersion parameter as a function of temperature predicted by using the 
exponential trapping energy distribution. 
If the linearity holds, the characteristic trapping energy $E_0$ can be estimated 
from the slope of the dispersion parameter against temperature. 
Such analysis has been performed using charge densities in bulk heterojunction
photovoltaic composites measured by light-induced electron spin resonance (LESR) techniques.~\cite{Seki13}

In the absence of the deep trap states expressed by the exponential trapping energy distribution, 
the rate of photovoltage decay as a function of light intensity obeys the power law with the exponent $1/2$. 
In this case, the exponent is independent of temperature. 
The change in the photovoltaic decay under continuous light irradiation with temperature 
indicates the presence of trap states with the distribution of trapping energies. 

In an experiment, 
the application of a weak light pulse gives rise to the multiexponential decay of photovoltage 
under the continuous irradiation of strong light.~\cite{Li} 
The reason for this is not clear but the multiexponential decay could be obtained in the MT model 
if there are separate regions described by independent trapping, detrapping, and recombination processes 
with different trapping energy distributions. 
In this case, the photovoltage decay can be described 
by the weighted sum of the exponential decay with the rate 
given by Eq. (\ref{eq:pvvsI}), where $\alpha$ is the dispersion parameter of each region. 
The weights could be independent of light intensity and temperature.

%


\begin{thebibliography}{99} %
\bibitem{Deibel}
C.~Deibel and V.~Dyakonov: Rep. Prog. Phys. {\bf 73} (2010) 096401. 

\bibitem{Ohkita2010}
J.~Guo, H.~Ohkita, H.~Benten, and S.~Ito: 
J. Am. Chem. Soc. {\bf 132} (2010) 9631 .

\bibitem{Nelson99} J.~Nelson: Phys. Rev. B {\bf 59} (1999) 15374. 

\bibitem{Nelson03} 
J.~Nelson: Phys. Rev. B {\bf 67} (2003) 155209.

\bibitem{Barzykin02} 
A.~V.~Barzykin and M.~Tachiya: J. Phys. Chem. B {\bf 106} (2002) 4356.

\bibitem{Barzykin04} 
A.~V.~Barzykin and M.~Tachiya: J. Phys. Chem. B {\bf 108} (2004) 8385.

\bibitem{Tachiya10} 
M.~Tachiya and K.~Seki: Phys. Rev. B {\bf 82} (2010) 085201.

\bibitem{Schmidlin} 
F.~W.~Schmidlin: Phys. Rev. B   {\bf 16} (1977) 2362.  

\bibitem{Rudenko1} 
A.~I.~Rudenko and V.~I.~Arkhipov: Philos. Mag. B {\bf 45} (1982) 177. 

\bibitem{Rudenko2} A.~I.~Rudenko and V.~I.~Arkhipov: Philos. Mag. B {\bf 45} (1982) 189.

\bibitem{Rudenko3} A.~I.~Rudenko and V.~I.~Arkhipov: Phillos. Mag. B {\bf 45} (1982) 209.  

 \bibitem{Zaban} A.~Zaban, M.~Greenshtein, and J.~Bisquert: ChemPhysChem {\bf 4} (2003) 859.

\bibitem{Foertig} 
A.~Foertig, A.~Baumann, D.~Rauh, V.~Dyakonov, and C.~Deibel: Appl. Phys. Lett. {\bf 95} (2009) 052104.
%
\bibitem{Montanari} 
I.~Montanari, A.~F.~Nogueira, J.~Nelson, J.~R.~Durrant, C.~Winder, M.~A.~Loi, N.~S.~Sariciftci, and C.~Brabec: Appl. Phys. Lett. {\bf 81} (2002) 3001.  


\bibitem{ShuttlePRB} 
C.~G.~Shuttle, B.~O'Regan, A.~M.~Ballantyne, J.~Nelson, D.~D.~C.~Bradley, and J.~R.~Durrant: Phys. Rev. B {\bf 78} (2008) 113201. 

\bibitem{Juska} 
G.~Ju\v{s}ka, K.~Genevicius, N.~Nekrasas, G.~Sliauzys, and G.~Dennler: Appl. Phys. Lett. {\bf 93} (2008) 143303.

\bibitem{Clarke} 
T.~M.~Clarke, F.~C.~Jamieson, and J.~R.~Durrant: J. Phys. Chem. C {\bf 113} (2009) 20934.

\bibitem{Nogueira}
A.~F.~Nogueira, I.~Montanari, J.~Nelson, J.~R.~Durrant, C.~Winder, N.~S.~Sariciftci, and C.~J.~Brabec: 
J. Phys. Chem. B {\bf 107} (2002) 1567. 

\bibitem{Ohkita}
S.~Yamamoto, H.~Ohkita, H.~Benten, and S.~Ito: 
J. Phys. Chem. C {\bf 116} (2012) 14804 . 

\bibitem{Ohkita2008}
H.~Ohkita, S.~Cook, Y.~Astuti, W.~Duffy, S.~Tierney, W.~Zhang, M.~Heeney, L.~McCulloch, J.~Nelson, D.~D.~C.~Bradley, and J.~R.~Durrant: 
J. Am. Chem. Soc. {\bf 130} (2008) 3030. 

\bibitem{Mozer}
T.~M.~Clarke, J.~Peet, P.~Denk, G.~Dennler, C.~Lungenschmied, and A.~J.~Mozer: 
Energy Environ. Sci. {\bf 5} (2012) 5241. 

\bibitem{Eng} 
M. ~P.~Eng, P.~R.~F.~Barnes, and J.~R.~Durrant: J. Phys. Chem. Lett. {\bf 1} (2010) 3096.

\bibitem{Naito} 
T.~Kobayashi, Y.~Terada, T.~Nagase, and H.~Naito: 
Appl. Phys. Express {\bf 4} (2011) 126602.

\bibitem{Seki13} 
K.~Seki, K.~Marumoto, and M.~Tachiya: Appl. Phys. Express {\bf 6} (2013) 051603.

\bibitem{Wurfel}
P. W\"{u}rfel: {\it Physics of Solar Cells} (Wiley-VCH, Weinheim, 2009) 2nd ed. p. 61 and p. 181.

\bibitem{Garcia-Belmonte} 
G.~Garcia-Belmonte and J.~Bisquert: Appl. Phys. Lett. {\bf 96} (2010) 113301.

\bibitem{Shuttle} 
C.~G.~Shuttle, B.~O'Regan, A.~M.~Ballantyne, J.~Nelson, 
D.~D.~C.~Bradley, J.~de~Mello, and J.~R.~Durrant: Appl. Phys. Lett. {\bf 92} (2008) 093311.

\bibitem{Li} 
Z.~Li , F.~Gao, N.~C.~Greenham, and C~R.~McNeill: 
Adv. Funct. Mater. {\bf 21} (2011) 1419.

\bibitem{Beiley}  Z.~M. Beiley, E.~T. Hoke, R.~Noriega, J.~Dacu\~{n}a, G.~F.~Burkhard,
J.~A.~Bartelt, A.~Salleo, M.~F.~Toney, and M.~D.~McGehee: Adv. Energy
Mater. {\bf 1} (2011) 954. 

\bibitem{Mark} 
P.~Mark and W.~Helfrich: J. Appl. Phys. {\bf 33} (1962) 205. 

\bibitem{Blom} 
P.~W.~M.~Blom, M.~J.~M.~de Jong, and J.~J.~M. Vleggaar: Appl. Phys. Lett.
{\bf 68} (1996) 3308.

\bibitem{Campbell} A.~J.~Campbell, D.~D.~C.~Bradley, and D.~G.~Lidzey: J. Appl. Phys. {\bf 82} (1997) 
6326.

\bibitem{Chiguvare} Z.~Chiguvare and V.~Dyakonov: Phys. Rev. B {\bf 70} (2004) 235207.

\bibitem{Street} R.~A.~Street: Phy. Rev. B {\bf 84} (2011) 075208. 

\bibitem{Abramowitz} 
M. Abramowitz and I. A. Stegun: {\it Handbook of Mathematical Functions} (Dover, New York, 1972) p. 556.

\bibitem{Mandoc} 
M.~M.~Mandoc, W.~Veurman, L.~J.~A.~Koster, B.~de~Boer, and P.~W.~M.~Blom: Adv. Funct. Mater. {\bf 17} (2007) 2167.

\bibitem{MandocAPL2007} 
M.~M.~Mandoc, F.~B.~Kooistra, J.~C.~Hummelen, B.~de~Boer, and P.~W.~M.~Blom:  Appl. Phys. Lett. {\bf 91} (2007) 263505.

\bibitem{Cowan}
S.~R. Cowan, A.~Roy, and A.~J.~Heeger:  Phys. Rev. B {\bf 82} (2010) 245207. 

\bibitem{Blakesley} J.~C.~Blakesley and D.~Neher: Phys. Rev. B {\bf 84} (2011) 075210.

\bibitem{Kirchartz} T.~Kirchartz, B.~E.~Pieters, J.~Kirkpatrick, U.~Rau, and J.~Nelson: 
Phys. Rev. B {\bf 83} (2011) 115209.

\bibitem{Koster05} 
L.~J.~A.~Koster, V.~D.~Mihailetchi, R.~Ramaker, and P.~W. ~M.~Blom: Appl. Phys. Lett. {\bf 86} (2005) 123509.

\bibitem{Deibel10} C.~Deibel, T.~Strobel, and V.~Dyakonov: Adv. Mater. {\bf 22} (2010) 4097.

\end{thebibliography}
\end{document}